\documentclass[12pt,a4paper]{article}

\usepackage{amsthm,amsmath,natbib,amssymb,amsfonts,bm, mathtools}
\usepackage{graphicx}
\usepackage{placeins}
\usepackage{epstopdf}
\usepackage{secdot}
\usepackage{longtable}
\usepackage{xcolor}
\usepackage{listings}
\usepackage{subfig}
\usepackage{arydshln}
\usepackage{bm}
\usepackage{upgreek}
\usepackage{url}
\usepackage{enumitem}
\usepackage[colorlinks, citecolor=blue, urlcolor=blue, linkcolor=blue]{hyperref}

\usepackage{geometry} % to change the page dimensions
\usepackage{sectsty} % RM added command to change section heading size - 23/8/17
\sectionfont{\fontsize{12}{16}\selectfont} % RM added command to change section heading size - 23/8/17
\subsectionfont{\fontsize{12}{16}\selectfont} % RM added command to change subsection heading size - 24/8/17

\newtheorem*{theorem*}{Theorem}

\theoremstyle{definition}

\theoremstyle{remark}

\DeclareMathAlphabet{\mathpzc}{OT1}{pzc}{m}{it}

\newcommand{\E}{\text{E}}

\makeatletter
\newcommand{\vast}{\bBigg@{4}}
\newcommand{\Vast}{\bBigg@{5}}
\makeatother

\newcounter{Dsystem}

\begin{document}
	
	\title{Adaptive-precision computation of custom Gauss quadrature for statistical applications}

\author{Paul Kabaila \\
\textsl{\small Department of Mathematical and Physical Sciences} \\
\textsl{\small La Trobe University, Bundoora, Victoria 3086, Australia}}
% \texttt{p.kabaila@latrobe.edu.au}

\date{}
	
\maketitle

\begin{center}	
\section*{Abstract}
\end{center}

\noindent An $n$-point Gauss quadrature rule
approximates the weighted integral of a function
by a weighted sum of $n$ evaluations of this function and is exact for polynomials of degree at most $2n-1$. Such rules can be highly accurate with relatively few 
evaluations of this function.
For weight functions associated with
classical orthogonal polynomials of a continuous variable (such as Legendre, Hermite and Laguerre), these quadrature rules are readily available.
We suppose that this is not the case, so that these rules must be custom-made. We present \textsf{CustomGaussQuadrature}, a \textsf{Julia} package that implements two approaches for computing the Gauss quadrature nodes and weights: the moment determinants method and the Stieltjes procedure. The principal contribution is an implementation of the ill-conditioned moment determinants method with adaptively chosen working precision. Comparisons between computations at increasing precisions provide practical error indicators used to target absolute and relative errors in the Gauss rule nodes and weights, respectively, of about $10^{-16}$. The package also adapts the working precision and number of auxiliary quadrature nodes used for the Stieltjes procedure. Numerical examples using scaled chi and Weibull probability density functions as weight functions show close agreement between the results obtained by the two approaches. The package is intended particularly for statistical applications in which a single custom Gauss rule is computed once and then reused to approximate many weighted integrals having the same weight function but different computationally expensive integrands.

\medskip
\noindent\textbf{Keywords:}
Gauss quadrature nodes and weights; Julia; moment determinants method;
Stieltjes procedure

	\newpage

\section{Introduction}\label{sec1}

In statistical applications, we often need to evaluate 
\begin{equation}
	\label{eqn_integral_to_be_evaluated}	
	\int_{-\infty}^{\infty}	g(x) \, f(x) \, dx,
\end{equation}
where $f$ is a specified nonnegative 
integrable 
weight function on $\mathbb{R}$. If $f$ is the probability density function (pdf) of a random variable $X$, \eqref{eqn_integral_to_be_evaluated} equals
$\E\big(g(X)\big)$. The Gauss quadrature approximation to this integral has the form
\begin{equation*}
	\sum_{i = 1}^n \lambda_{i} \, g(\tau_{i}),
\end{equation*}
where $\tau_1, \dots, \tau_n$ are called the nodes and 
$\lambda_1, \dots, \lambda_n$ are called the corresponding weights.
The dependence of these nodes and weights on $n$ is implicit. This approximation is exact when $g$ is a 
polynomial of degree at most $2n - 1$. Together, these nodes and weights form the Gauss 
%quadrature 
rule with $n$ nodes. 
We suppose that $f$ is positive almost everywhere on a nondegenerate interval $I$, and zero almost everywhere outside it. We call the smallest closed interval containing $I$ the support interval.
 We also assume that
\begin{equation*}
	\int_{-\infty}^{\infty}	|x|^{2 n -1} \, f(x) \, dx < \infty.
\end{equation*}
These assumptions guarantee the existence of the $n$-node Gauss rule.

When $f$ leads to Gauss 
%quadrature 
rules with nodes that are the roots of 
classical orthogonal polynomials of a continuous variable (such as Legendre, Hermite and Laguerre) then these rules are readily available; this is the classical case.
Otherwise, the Gauss rule must be computed from scratch. 
To illustrate, consider the following non-classical weight function.
% , 
% referred to below as the \emph{Scaled chi pdf example}.
\smallskip

\noindent \textbf{Scaled chi pdf weight function}:
Let the weight function $f$ be the pdf of $X = R / m^{1/2}$, where $R$ has a $\chi_m$ distribution (i.e. $R^2$ has a $\chi_m^2$ distribution).
Here, the positive integer $m$ is the number of degrees of freedom. This pdf is given by 
\begin{equation*}
	%	\label{eqn_pdf_of_Example}
	\begin{split}
		f(x) =
		\begin{cases}
			\dfrac{m^{m/2}}{\Gamma(m/2) \, 2^{(m/2) - 1}} \; x^{m-1} \, \exp\big(- m \, x^2 /2\big) &\text{for } x > 0
			\\
			0 &\text{otherwise}.
		\end{cases}	
	\end{split}
\end{equation*}

\smallskip

\noindent Integrals of the form \eqref{eqn_integral_to_be_evaluated} with this weight function
arise in simultaneous statistical inference  
and the selection and ranking of populations 
(\citeauthor{Miller1981}
\citeyear{Miller1981}; 
\citeauthor{HochbergTamhane1987} \citeyear{HochbergTamhane1987};
\citeauthor{GuptaPanchapakesan2002} \citeyear{GuptaPanchapakesan2002})
and in the computation of central and non-central (Kshirsagar definition)
multivariate $t$ probabilities
(\citeauthor{DunnettSobel} \citeyear{DunnettSobel}; 
\citeauthor{Dunnett} \citeyear{Dunnett}; 
\citeauthor{GenzBretz} \citeyear{GenzBretz}),
when $g(x)$ is computed by the method of \cite{MiwaHayterKuriki2003}.

\cite{AbeysekeraKabaila2017}, \cite{KabailaGiri2009JSPI},
\cite{KabailaGiri2013} and \cite{KabailaTissera2014} 
used numerical nonlinear constrained optimization to
construct confidence regions, with guaranteed minimum coverage, that utilize uncertain prior information. \cite{KabailaWelshAbeysekera2016}, \cite{KabailaWelshWijethunga2020_JSPI} 
and \cite{KabailaPerera2023} numerically assess the performance, in terms of coverage probability and expected length, of some frequentist model averaged confidence intervals. 
These constructions and performance assessments require the evaluation of integrals of the form \eqref{eqn_integral_to_be_evaluated} for a \emph{Scaled chi pdf weight function}
 $f$, for given $m$, for tens of thousands or even hundreds of thousands of different functions $g$, where the evaluation of $g(x)$ for any given $x$ is computationally very expensive. A fixed quadrature rule can be computed once and then reused throughout these evaluations. In this setting, obtaining high accuracy with relatively few evaluations of $g(x)$ is particularly valuable. 
\cite{KabailaRanathunga2021} computed these integrals using the transformation (2.6) of \cite{Mori1988} followed by the trapezoidal rule, but this approach is much
less accurate than a custom-made Gauss rule with the same number of function evaluations.

The general-purpose \textsf{Julia} package \textsf{CustomGaussQuadrature}
(\citeauthor{Kabaila_CustomGaussQuadrature_2026},
\citeyear{Kabaila_CustomGaussQuadrature_2026})
was developed for this repeated-use setting. 
The stopping criteria are designed to produce Gauss rule nodes and weights with absolute and relative errors, respectively, of about $10^{-16}$, although this accuracy is not guaranteed.
Once computed, this Gauss rule can be used repeatedly in \texttt{Float64} calculations involving many computationally expensive functions $g$. 
The Gauss rule is computed in two steps as follows.

\smallskip

\noindent\makebox[3em][l]{\textbf{Step 1}:} \ \ \hangindent=1.5em Compute the recursion coefficients in the three-term recurrence relation
(described in Section~2). 
In the classical case, simple formulae for the recursion coefficients make this step trivial; in the non-classical case, it is the main difficulty. 
\cite{Gautschi2004} describes several methods 
for this computation with widely varying complexity and sensitivity to roundoff errors. 
We
implement two methods: moment determinants and the
Stieltjes procedure. 

\smallskip

\noindent\makebox[3em][l]{\textbf{Step 2}:} \ \ \ \hangindent=1.5em
Use these recursion coefficients to compute the Gauss rule.
This step is the same for both classical and non-classical cases.
% The map from the recursion coefficients to the Gauss quadrature nodes weights is well-conditioned
% and the map from the recursion coefficients to the Gauss quadrature weights is reasonably 
% well-conditioned under a simple condition (see
% Section \ref{sect_step2_julia_code} for details). 
We carry out this step by computing the eigenvalues and eigenvectors of the symmetric
tridiagonal matrix defined by the recursion coefficients with a command from 
the package \textsf{GenericLinearAlgebra}.

\smallskip

Although the moment determinants method is severely ill-conditioned
(Gautschi, \citeyear{Gautschi1968}, \citeyear{Gautschi1983}, 
\citeyear{Gautschi1994}, \citeyear{Gautschi2004}),
this limitation can be overcome by the use of high-precision arithmetic, see e.g. \cite{Gautschi1983}.
%and p.219 of  \cite{Milanovic2015}. 
Our main contribution is to implement the moment determinants method with an adaptive choice of number of bits of precision \texttt{b} of 
%\textsf{Julia}'s 
\texttt{BigFloat}  
arithmetic, 
so that the computed recursion coefficients are sufficiently accurate.

To check whether the chosen precision \texttt{b} is sufficiently large,
we use the following simple method. Suppose that $c_1$ and $c_2$ are numerical approximations to the same quantity computed using the same 
%\textsf{Julia} 
function
that implements an ill-conditioned numerical method, 
using \texttt{BigFloat} arithmetic with number of bits of precision \texttt{b1} and \texttt{b2}, respectively, where \texttt{b2} is substantially larger than \texttt{b1}.
The approximation $c_2$ is therefore believed to be much more accurate than $c_1$. We take $c_2$ as our final approximation and 
use $|c_1-c_2|$ as a practical error indicator (although not a rigorous upper bound). 

%This is facilitated by \textsf{Julia}'s support for type-generic numerical programming, allowing the same function definitions to be used with \texttt{Float64}, \texttt{Double64}, and \texttt{BigFloat} arithmetic.

The moment determinants method requires a formula 
for the $s$'th moment 
\begin{equation*}
	%	\label{eq_moments_as_integrals}	
	\mu_s
	= \int_{-\infty}^{\infty}	x^s \, f(x) \, dx
\end{equation*}
for all nonnegative integers $s \le 2n-1$. 
The Stieltjes procedure requires instead a formula for the weight function $f$.
However, it also requires the choice of an auxiliary quadrature rule with $r$ nodes, where $r$ needs to be chosen adaptively.
For many weight functions, one needs to take account of the special nature of the weight function to avoid excessively large values of $r$.
The Stieltjes procedure may also be ill-conditioned,
suggesting that the arithmetic precision should also be adaptively chosen. This makes the 
Stieltjes procedure more demanding to apply than might first seem the case.

Section~2 describes the three-term recurrence relation and the two Step~1 methods. Step~2 is described in Section~3. Sections~4 and~5 present the implementations of Steps~1 and~2 in code. Section~6 presents two numerical examples comparing the moment determinants method and the Stieltjes procedure. Section~7 briefly reviews other \textsf{Julia} packages that use the Stieltjes procedure, and Section~8 provides a discussion. Appendices~A and~B give further implementation details for the moment determinants method and the Stieltjes procedure, respectively.

\section{Step 1 using either moment determinants or the Stieltjes procedure}
\label{sect_step1_moments_or_stieltjes}

We first describe the three-term recurrence relation. Then we describe the computation of the recursion coefficients using either moment determinants or the Stieltjes procedure. 

\subsection{The three-term recurrence relation}

For any function $u: \mathbb{R} \rightarrow  \mathbb{R}$, the norm of $u$ is defined to be the square root of
\begin{equation*}
	\int_{-\infty}^{\infty} u^2(x) \,f(x) \, dx,
\end{equation*}
where $f$ is the weight function, provided that this integral exists. This norm is denoted by  $\| u \|$.
If two functions 
$u: \mathbb{R} \rightarrow  \mathbb{R}$
and $v: \mathbb{R} \rightarrow  \mathbb{R}$
satisfy $\| u \| < \infty$ and $\| v \| < \infty$
then we define $(u, v)$,
the inner product of these two functions, to be 
\begin{equation}
	\label{eqn_inner_prod_fns}
	\int_{-\infty}^{\infty}u(x) \, v(x) \, f(x) \, dx.
\end{equation}
If $(u,v) = 0$ then the functions $u$ and $v$ are said to be orthogonal. 

Henceforth, we consider only polynomials whose coefficients are real numbers. A polynomial in $x$ of degree $n$ is said to be monic if the coefficient of $x^n$ is 1. Let $\pi_k$ denote a monic polynomial of degree $k$. The monic polynomials $\pi_0, \pi_1, \dots, \pi_{n-1}$ are called monic orthogonal polynomials with respect to the weight function $f$ if 
\begin{align*}
	(\pi_k, \pi_{\ell}) &= 0 \ \ \text{for all} \ k \ne \ell, \ \text{where} \ k, \ell \in \{0, 1, \dots, n-1\}
	\\
	\| \pi_k \| &> 0 \ \ \text{for} \ k=0, 1, \dots, n-1.
\end{align*}
The monic polynomial $\pi_n$ is defined by the conditions
\begin{equation*}
	\int_{-\infty}^{\infty} \pi_n(x)\,\pi_k(x)\,f(x)\,dx=0
	\quad \text{for} \quad k=0,1,\dots,n-1.
\end{equation*}
The Gauss quadrature nodes $\tau_1, \dots, \tau_n$ are the $n$ distinct roots of the polynomial $\pi_n$. 

The polynomials $\pi_0, \pi_1, \dots, \pi_n$ satisfy the following three-term recurrence relation, see e.g.
Theorem 1.27 on p.10 of \cite{Gautschi2004}. Let 
$\pi_{-1} \equiv 0$ and $\pi_0 \equiv 1$. Then 
\begin{equation}
	\label{eqn_3_term_recurrence_1}	
	\pi_{k+1}(x) = 	\big(x - \alpha_k\big) \, \pi_k(x) - \beta_k \, \pi_{k-1}(x) \ \ \text{for} \ k = 0, 1, \dots, n-1,
\end{equation}
where 
\begin{align}
	\label{eqn_3_term_recurrence_2}		
	\alpha_k 
	= \frac{\big(x \, \pi_k, \pi_k\big)}{\big(\pi_k, \pi_k\big)}
	\quad (k = 0, 1, \dots, n-1) \quad \text{and} \quad
	\beta_k 
	= \frac{\big(\pi_k, \pi_k\big)}{\big(\pi_{k-1}, \pi_{k-1}\big)}
	\quad (k = 1, 2, \dots, n-1).
\end{align}
The proof of this recurrence relation is based on Gram-Schmidt orthogonalization.

\subsection{Computation of the recursion coefficients in the three-term recurrence relation using moment determinants}
\label{sect_recursion_coeffs_moment_dets}

The recursion coefficients $\alpha_0, \alpha_1, \dots, \alpha_{n-1}$ and 
$\beta_1, \beta_2, \dots, \beta_{n-1}$ may be computed, in terms of determinants of matrices whose entries are moments, as follows. Let $\Delta_{\ell}$ denote the Hankel determinant of order ${\ell}$ in the moments $\mu_k$ given by 
\begin{align*}
	\Delta_0 = 1, \quad 
	\Delta_{\ell} 
	= \begin{vmatrix}
		\mu_0 & \mu_1 & \hdots & \mu_{{\ell}-1}\\ 
		\mu_1 & \mu_2 & \hdots & \mu_{\ell}\\
		\vdots & \vdots &      & \vdots \\
		\mu_{{\ell}-1} & \mu_{\ell} & \hdots & \mu_{2{\ell}-2} 
	\end{vmatrix},	
	\quad {\ell} = 1, 2, \dots, n.
\end{align*}
Also define
\begin{align*}
	\Delta_0^{\prime} = 0, \quad 
	\Delta_1^{\prime} = \mu_1, \quad 
	\Delta_{\ell}^{\prime} 
	= \begin{vmatrix}
		\mu_0 & \mu_1 & \hdots & \mu_{{\ell}-2} & \mu_{\ell} \\ 
		\mu_1 & \mu_2 & \hdots & \mu_{{\ell}-1} & \mu_{{\ell}+1}\\
		\vdots & \vdots &      & \vdots & \vdots \\
		\mu_{{\ell}-1} & \mu_{\ell} & \hdots & \mu_{2{\ell}-3} & \mu_{2{\ell}-1}
	\end{vmatrix},
	\quad {\ell} = 2, 3, \dots, n.
\end{align*}
In other words, $\Delta_{\ell}^{\prime}$ is the Hankel determinant 
$\Delta_{{\ell}+1}$ with the second-last column and the last row removed. According to Theorem 2.2 on p.54 of 
\cite{Gautschi2004}, 
\begin{equation}
	\label{eqn_alphak_moment_det}	
	\alpha_k 
	= \frac{\Delta_{k+1}^{\prime}}{\Delta_{k+1}}
	- \frac{\Delta_{k}^{\prime}}{\Delta_{k}} \quad 
	(k = 0, 1, 2, \dots, n-1)	
\end{equation}
and 
\begin{align}
	\label{eqn_betak_moment_det}
	\beta_k 
	&= \frac{\Delta_{k+1} \, \Delta_{k-1}}{\Delta_k^2} \quad 
	(k = 1, 2, 3, \dots, n-1).
\end{align}

\subsection{Computation of the recursion coefficients in the three-term recurrence relation using the Stieltjes procedure}
\label{sect_recursion_coeffs_stieltjes_procedure}

The Stieltjes procedure, given in subsections 2.2.2 and 2.2.3 of \cite{Gautschi2004}, applies the three-term recurrence relation \eqref{eqn_3_term_recurrence_1} and \eqref{eqn_3_term_recurrence_2}	iteratively through the following sequence:
\begin{equation}
	\label{eqn:stieltjes procedure outline}
	\big\{\pi_{-1}, \pi_0 \big\} \rightarrow
	\big\{\alpha_0, \pi_1 \big\} \rightarrow
	\big(\pi_1, \pi_1\big) \rightarrow
	\beta_1  \rightarrow
	\big\{\alpha_1, \pi_2 \big\} \rightarrow
	\big(\pi_2, \pi_2\big) \rightarrow
	\beta_2  \rightarrow \dots.
\end{equation}
The inner products used in the computation of the recursion coefficients are found using a high-quality quadrature rule with $r$ nodes. We call this the auxiliary quadrature rule. This is a discretization method that is expected to lead to
approximations to the recursion coefficients $\alpha_0, \alpha_1, \dots, \alpha_{n-1}$ and $\beta_1, \beta_2, \dots, \beta_{n-1}$
that converge to their exact values as $r \rightarrow \infty$.
Roughly speaking, the Stieltjes procedure may be viewed as a method for 
converting one high-quality quadrature rule (the auxiliary quadrature rule) into another high-quality quadrature rule (the Gauss rule).

\section{Step 2 using the eigenvalues and eigenvectors of the Jacobi matrix}
\label{sect_step2_jacobi_matrix}

For $n=1$, $\lambda_1 = \mu_0$ and $\tau_1 = \alpha_0$.
For $n \ge 2$, define the $n \times n$ Jacobi matrix 
\begin{equation*}
	J_n = \left[
	\begin{array}{cccccc}
		\alpha_0 & \sqrt{\beta_1} & 0 & 0 & \hdots & 0 \\
		\sqrt{\beta_1} & \alpha_1 & \sqrt{\beta_2} & 0 & \ddots & \vdots \\
		0 & \sqrt{\beta_2} & \alpha_2 & \sqrt{\beta_3} & 0  & 0\\
		0 & \ddots & \ddots & \ddots & \ddots & 0\\
		\vdots & \ddots & & \sqrt{\beta_{n-2}} & \alpha_{n-2} & \sqrt{\beta_{n-1}} \\
		0 & \hdots & 0 & 0 & \sqrt{\beta_{n-1}} & \alpha_{n-1} 
	\end{array}	
	\right].
\end{equation*}
The nodes $\tau_1, \dots, \tau_n$ are the eigenvalues of $J_n$, in increasing order. Let $\bm{x}_i$ denote an eigenvector corresponding to the eigenvalue $\tau_i$. The weight
\begin{equation*}
	\lambda_i 
	= \frac{\mu_0 \, (\text{first component of }\bm{x}_i)^2}
	{(\bm{x}_i, \bm{x}_i)}.
\end{equation*}
%
%where 
%
%\begin{equation*}
%\mu_0 =	\int_{-\infty}^{\infty}	f(x) \, dx.
%\end{equation*}
%
%see e.g. Theorem 1.3.1 of \cite{Gautschi2004}.
The matrix $J_n$ is tridiagonal; its nonzero elements are only on the subdiagonal, diagonal and superdiagonal. It is also symmetric.

\section{Implementation of the methods for Step 1 in code}

We describe the implementation in \textsf{Julia} code of Step~1 using moment determinants,
as described in Section \ref{sect_recursion_coeffs_moment_dets}, and
using the Stieltjes procedure, as described in Section \ref{sect_recursion_coeffs_stieltjes_procedure}.

\subsection{Implementation of the moment determinants method  
	for Step 1 in \textsf{Julia} code}
\label{sect_meth_moment_dets_step1_julia}	

The map from the vector 
$\big(\mu_0, \mu_1, \dots, \mu_{2n-1} \big)$ of moments to the vector $\big(\alpha_0, \dots, \alpha_{n-1},\newline  \beta_1, \dots, \beta_{n-1}\big)$ of recursion coefficients is severely ill-conditioned (Gautschi, \citeyear{Gautschi1968}, \citeyear{Gautschi1983}, 
\citeyear{Gautschi1994}, \citeyear{Gautschi2004}). However, it has long been recognized that this limitation can be overcome by the use of high-precision arithmetic, see e.g. \cite{Gautschi1983},
%and p.219 of  \cite{Milanovic2015}, 
when applying 
the moment determinants method.

We compute approximations to the vectors
\begin{equation*}
    \big(\alpha_0, \dots, \alpha_{n-1} \big)
    \quad \text{and} \quad
    \big(\sqrt{\beta_1}, \dots, \sqrt{\beta_{n-1}}\big)
\end{equation*}
for which the successive-precision differences satisfy maximum absolute and maximum relative stopping tolerances, respectively, of $10^{-18}$. 
The script \path{gauss_quad_moment_dets_scr.jl} carries out both Steps 1 and 2, where Step 1 uses moment determinants. The part of this script that carries out Step 1 is described in detail in Appendix A. Let 
$\big(\alpha_0(\texttt{b}), \alpha_1(\texttt{b}), \dots, \alpha_{n-1}(\texttt{b})\big)$, 
$\big(\beta_1(\texttt{b}), \beta_2(\texttt{b}), \dots,\beta_{n-1}(\texttt{b}) \big)$ and $\mu_0(\texttt{b})$
denote the approximations to 
$\big(\alpha_0, \alpha_1, \dots, \alpha_{n-1}\big)$, 
$\big(\beta_1, \beta_2, \dots,\beta_{n-1} \big)$ and $\mu_0$,
respectively, obtained using  \texttt{BigFloat} arithmetic with 
\texttt{b} bits of precision. 
The output of this part is
\begin{align*}
    &\big(a_1(\texttt{b}_{\rm final}), a_2(\texttt{b}_{\rm final}), \dots,
    a_n(\texttt{b}_{\rm final})\big), \\
    &\big(b_1(\texttt{b}_{\rm final}), b_2(\texttt{b}_{\rm final}), \dots,
    b_{n-1}(\texttt{b}_{\rm final}) \big), \\
    &\mu_0(\texttt{b}_{\rm final}) \quad \text{and} \quad \texttt{b}_{\rm final},
\end{align*}
where $\texttt{b}_{\rm final}$ denotes the final choice for \texttt{b}.

\subsection{Implementation of the Stieltjes procedure for Step 1 in 
	 code}
\label{sect: Implement Stieltjes for Step 1}	

To implement the Stieltjes procedure, described in Section \ref{sect_recursion_coeffs_stieltjes_procedure}, 
we need to approximate the inner product $(u, v)$ of two functions $u$ and $v$ given by 
\eqref{eqn_inner_prod_fns}. Let the function $g = u \, v$. In Appendix B we describe the method used 
to compute the $w_i$'s and $x_i$'s in the $r$-node discrete approximation  
\begin{equation}
	\label{eqn_discrete_approx_inner_prod_fns}
	\sum_{i =1}^r w_i \, g(x_i)
\end{equation}
to $(u,v)$. We call this the auxiliary quadrature rule.

We suppose that $f$ is positive almost everywhere on an interval with lower and upper endpoints $a$ and $b$, respectively, and zero almost everywhere outside this interval. Here $-\infty \le a < b \le \infty$. The inner product of the functions $u$ and $v$ is therefore 
\begin{equation*}
	\int_a^b g(x) \, f(x) \, dx,
\end{equation*}
where $g = u \, v$.

To compute an approximation to this integral, 
we use the ``general purpose'' initial transformation described
on p.94 of \cite{Gautschi2004}. In other words, we transform the  
support interval with lower and upper endpoints $a$ and $b$, respectively, to the interval with lower and upper endpoints $-1$ and $1$, respectively, using the 
transformation
\begin{equation}
	\label{eqn_initial_transf}
	\int_a^b g(x) \, f(x) \, dx	 
	= \int_{-1}^1 g\big(\varphi(y)\big) \, f\big(\varphi(y)\big) \, \varphi'(y) \, dy
\end{equation}
where
\begin{equation*}
	\varphi(y) = 
	\begin{cases}
		(1/2) (b - a) y + (1/2) (b + a) &\text{if } -\infty < a < b < \infty
		\\
		b - (1 - y) / (1 + y) &\text{if } -\infty = a < b < \infty
		\\
		a + (1 + y) / (1 - y) &\text{if } -\infty < a < b = \infty
		\\
		y / (1 - y^2)  &\text{if } -\infty = a < b = \infty.
	\end{cases}	
\end{equation*}
A high-quality quadrature rule then provides a discrete approximation to the right-hand side of \eqref{eqn_initial_transf}.
\cite{Gautschi2004} used a Fej\'er quadrature rule.
Instead, we use Gauss--Legendre quadrature.

We compute approximations to the vectors
\begin{equation*}
    \big(\alpha_0, \alpha_1, \dots, \alpha_{n-1}\big)
    \quad \text{and} \quad
    \big(\sqrt{\beta_1}, \sqrt{\beta_2}, \dots, \sqrt{\beta_{n-1}} \big)
\end{equation*}
for which the successive-$r$ differences satisfy maximum absolute and maximum relative stopping tolerances, respectively, of $10^{-17}$. This is less stringent than when these vectors are computed using moment determinants because the value of $r$ required for sufficiently accurate results can be remarkably large.

In the comments for his subroutine \texttt{qgp}
(available at \url{cs.purdue.edu/archives/2002/wxg/codes}) which uses the initial transformation \eqref{eqn_initial_transf},
Gautschi states of this method that
``It takes no account of the special nature of the weight function involved and hence may result in slow convergence of the discretization procedure. This routine, therefore, 
should be used only as a last resort, when no better, more natural 
discretization can be found.''

For given $n$, the computation of $\alpha_{n-1}$ requires the computation of the right-hand side of \eqref{eqn_initial_transf} with $g$ a polynomial of degree $2 n - 1$. 
To illustrate this potential difficulty of this computation consider
the computation of the right-hand side of \eqref{eqn_initial_transf} for $g(x) \equiv x^s$ for $s = 2n -1$. In other words, consider the computation of 
\begin{equation}
	\label{eqn:illustrate Gautschi's statement}
	\int_{-1}^1 \varphi^s(y) \, f\big(\varphi(y)\big) \, \varphi'(y) \, dy.
\end{equation}
Suppose that the weight function is the following.
%,
%referred to below as the \emph{Weibull pdf example}.

\smallskip

\noindent \textbf{Weibull pdf weight function}:
Let $f$ be the Weibull pdf with 
shape parameter $k > 0$ and scale parameter 1, so that
\begin{equation*}
	f(x) =
	\begin{cases}
		k \, x^{k - 1} \exp(-x^{k}) &\ \text{for} \ \ x > 0
		\\
		0  &\ \text{otherwise.}
	\end{cases}
\end{equation*}

\smallskip

\noindent  Also suppose that  $n = 10$, so that $s = 19$. 
% This graph is obtained using test\transformed_weibull_scr.jl
The graph of the integrand of  \eqref{eqn:illustrate Gautschi's statement} has a single peak which approaches 1 and has width approaching 0, as the parameter $k$ approaches 0. This implies that the value of $r$
%, the number of nodes in the discrete approximation (4), 
required for 
sufficiently accurate results from the Stieltjes procedure
increases rapidly as $k$ approaches 0.

\section{Implementation of the method for Step 2 in code}
\label{sect_step2_julia_code}

The Gauss rule nodes and weights are found by computing the eigenvalues and eigenvectors of the symmetric tridiagonal $n \times n$ Jacobi matrix $J_n$. Real symmetric matrices are Hermitian. On pp.~42--43, \cite{Stewart_MatrixAlgorithms_Eigensystems_2001} states that ``the eigenvalues of a Hermitian matrix are perfectly conditioned---that is, a perturbation in the matrix makes a perturbation that is no larger in the eigenvalues.'' Thus, with respect to perturbations in $J_n$, the Gauss rule nodes are perfectly conditioned in an absolute normwise sense.

The result (3.28) of \cite{Stewart_MatrixAlgorithms_Eigensystems_2001} implies that a unit eigenvector corresponding to a simple eigenvalue of a Hermitian matrix is well-conditioned when that eigenvalue is well separated from the other eigenvalues. Consequently, the computation of the Gauss rule weights is generally well-conditioned provided that
\begin{equation*}
	\frac{1}{\min\big(|\tau_2 - \tau_1|, \dots, |\tau_n - \tau_{n-1}|\big)}
\end{equation*}
is not too large. A caveat is that a small absolute error in an eigenvector component can produce a larger relative error when the corresponding quadrature weight is very small.

We use the package \textsf{GenericLinearAlgebra}, together with \texttt{Double64} floating-point arithmetic (essentially quadruple precision) from the package \textsf{DoubleFloats}, to compute the eigenvalues and eigenvectors of the Jacobi matrix $J_n$. These are then converted to the Gauss rule nodes and weights, as described in Section~\ref{sect_step2_jacobi_matrix}. The use of \texttt{Double64} arithmetic is intended to make the numerical error introduced in Step~2 negligible relative to that remaining after Step~1.

For Step 1 carried out using moment determinants, the stopping tolerance is $10^{-18}$ for the maximum absolute and maximum relative differences between successive computations of the two vectors, respectively. For Step 1 carried out using the Stieltjes procedure, the corresponding stopping tolerance is $10^{-17}$. These tolerances are intended to provide plausible indications of the accuracy of the resulting Gauss quadrature nodes and weights.

Once we have computed
\begin{equation*}
	\big(\alpha_0, \alpha_1, \dots, \alpha_{n-1}\big)
	\qquad \text{and} \qquad
	\big(\sqrt{\beta_1}, \sqrt{\beta_2}, \dots, \sqrt{\beta_{n-1}}\big),
\end{equation*}
we can easily compute the Gauss rules with $1, 2, \dots, n$ nodes. 
%Our \textsf{Julia} 
The code also provides the option of computing all of these Gauss rule nodes and weights.

%We assume that this package 
%implements a stable algorithm for this computation. 

\section{Two numerical examples}

The next two subsections describe the results of computing
the Gauss rule using moment determinants and the Stieltjes procedure
for two examples: the \emph{Scaled chi pdf weight function} and the \emph{Weibull pdf weight function}. To install 
and load \textsf{CustomGaussQuadrature}, use
\begin{lstlisting}[basicstyle=\ttfamily\small]
	using Pkg
	Pkg.add("CustomGaussQuadrature")
	using CustomGaussQuadrature
\end{lstlisting}
Similarly, install and load \textsf{SpecialFunctions} using
\begin{lstlisting}[basicstyle=\ttfamily\small]
	Pkg.add("SpecialFunctions")
	using SpecialFunctions
\end{lstlisting}
This package will be needed for the computation of the gamma function.

The weight function is specified by the array \texttt{which\_f}. 
Its first
component is a string naming the weight function, its second component is a
two-vector giving the support interval, and its optional third component gives
the parameter or parameters. Exact quantities such as integers and
\texttt{Inf} or \texttt{-Inf} should be entered as ordinary Julia values. 
For example, the scaled chi pdf weight function, with $m = 160$, is specified by 
\begin{lstlisting}[basicstyle=\ttfamily\small]
	which_f = ["scaled chi pdf", [0, Inf], 160]
\end{lstlisting}
Finite non-integer numbers should be
entered as strings. For example, the Weibull pdf weight function, with $k=3.1$, is specified by
\begin{lstlisting}[basicstyle=\ttfamily\small]
	which_f = ["weibull pdf", [0, Inf], "3.1"]
\end{lstlisting}
Use \texttt{"3.1"} rather than \texttt{3.1}. The adaptive-precision computations are carried out such that, for any specified precision, the most accurate representation of 3.1 is used.

\subsection{Computation of the Gauss rule using moment determinants}
\label{sect: Gauss rule using moment determinants example}

To apply the moment determinants method, we require a formula for the $s$'th moment $\mu_s$ 
for all nonnegative integers $s \le 2 n - 1$ that can be evaluated using  \texttt{BigFloat}
arithmetic. For the Step 1 computations, \texttt{nbits} is the adaptively-chosen \texttt{BigFloat} precision. The outputs from the function \texttt{custom\_gauss\_quad\_all\_fn} include \texttt{nodes} and
\texttt{weights} which are \texttt{Double64} floating-point numbers (essentially quadruple precision).

\medskip

\noindent {\textbf{Example 1}} \ \ 
\textbf{Scaled chi pdf weight function with $\boldsymbol{m = 160}$ and $\boldsymbol{n = 3}$}

\smallskip

\noindent 
The values of $m$ and $n$ are
 taken from the Cholesterol-data illustration of \cite{KabailaPerera2023}.
For \emph{Scaled chi pdf weight function}
\begin{equation*}
	\mu_s =
	\left(\frac{2}{m} \right)^{s/2}
	\frac{\Gamma\big((s+m)/2\big)}{\Gamma(m/2)},
\end{equation*}
for $s = 0, 1, 2, \dots$. This formula is already implemented in
\texttt{moment\_stored\_fn}. 
The Gauss rule is computed, via moment determinants, using
\begin{lstlisting}[basicstyle=\ttfamily\small]
which_f = ["scaled chi pdf", [0, Inf], 160]		
n = 3

nodes, weights, nbits = 
	custom_gauss_quad_all_fn(moment_stored_fn, which_f, n)

nodes_momentdets_ex1 = nodes
weights_momentdets_ex1 = weights
println("nbits = ", nbits, "\n")
\end{lstlisting}

\noindent {\textbf{Example 2}} \ \ 
\textbf{Weibull pdf weight function with $\boldsymbol{k = 3.1}$ and $\boldsymbol{n=6}$
	and $\boldsymbol{n=14}$} 

\smallskip

\noindent For the \emph{Weibull pdf weight function}
\begin{equation*}
	\mu_s = \Gamma\left(1 + \frac{s}{k}\right),
\end{equation*}
for $s = 0, 1, 2, \dots$. 
Specify the function 
for computing $\mu_s$ using
\begin{lstlisting}[basicstyle=\ttfamily\small,breaklines=true,breakatwhitespace=false,columns=fullflexible]
function moment_weibull_pdf(::Type{T}, which_f, s::Integer) where {T<:AbstractFloat}
    @assert which_f[1] == "weibull pdf"
    T_k = CustomGaussQuadrature.materialize_scalar_spec_fn(T, which_f[3])
    @assert T_k > zero(T)
    @assert s >= 0
    if s == 0
        return one(T)
    end
    gamma(one(T) + convert(T, s) / T_k)
end
\end{lstlisting}

The Gauss rule is computed, via moment determinants, using
\begin{lstlisting}[basicstyle=\ttfamily\small]
which_f = ["weibull pdf", [0, Inf], "3.1"]	
n = 6

nodes, weights, nbits = 
	custom_gauss_quad_all_fn(moment_weibull_pdf, which_f, n)

nodes_momentdets_ex2 =  nodes
weights_momentdets_ex2 = weights
println("nbits = ", nbits, "\n")
\end{lstlisting}
This code was also run with $n=6$ replaced by $n=14$.

\subsection{Computation of the Gauss rule using the Stieltjes procedure}
\label{sect: Gauss rule using Stieltjes procedure example}

To compute the Gauss rule using the Stieltjes procedure, the user needs to supply a function for computing $\log(f(x))$ on the support of $f$, together with the value of $\mu_0$, as illustrated by the following two examples. For a weight function that is a pdf, $\mu_0 = 1$.
The Stieltjes scripts create the outputs \path{nodes_stieltjes},
\path{weights_stieltjes}, \path{a_vec_stieltjes},
\path{b_vec_stieltjes}, \path{nbits_stieltjes} and \texttt{r}
in the caller. Here, \path{nbits_stieltjes} and \texttt{r} are the adaptively-chosen \texttt{BigFloat} precision and number of auxiliary quadrature nodes, respectively. Also, \path{nodes_stieltjes} and
\path{weights_stieltjes} are \texttt{Double64} floating-point numbers (essentially quadruple precision).

\medskip

\noindent {\textbf{Example 1}} \ \ 
\textbf{Scaled chi pdf weight function with $\boldsymbol{m = 160}$ and $\boldsymbol{n = 3}$}

\smallskip

\noindent For the \emph{Scaled chi pdf weight function}, the formula for the log-weight function is already 
included in \texttt{stieltjes\_lnf\_stored\_scr.jl}. The Gauss rule is computed, via the Stieltjes procedure, using
\begin{lstlisting}[basicstyle=\ttfamily\small,breaklines=true,breakatwhitespace=false,columns=fullflexible]
which_f = ["scaled chi pdf", [0, Inf], 160]	
n = 3

pkg_dir = dirname(dirname(pathof(CustomGaussQuadrature)))
include(joinpath(pkg_dir, "src", "stieltjes_lnf_stored_scr.jl"))

nodes_stieltjes_ex1 = nodes_stieltjes
weights_stieltjes_ex1 = weights_stieltjes
println("nbits_stieltjes =", nbits_stieltjes)
println("r = ", r)
\end{lstlisting}

\smallskip

\noindent {\textbf{Example 2}} \ \ 
\textbf{Weibull pdf weight function with $\boldsymbol{k = 3.1}$ and $\boldsymbol{n=6}$
and $\boldsymbol{n=14}$}

\smallskip

\noindent 
For the \emph{Weibull pdf weight function},
\begin{equation*}
	\log(f(x)) = \log(k) + (k - 1)\log(x) - x^k \quad \text{for} \quad x > 0.
\end{equation*}
We specify the function for computing $\log(f(x))$ for $x > 0$ using
\begin{lstlisting}[basicstyle=\ttfamily\small,breaklines=true,breakatwhitespace=false,columns=fullflexible]
function lnf_weibull_pdf(::Type{T}, which_f, x::AbstractFloat) where {T<:AbstractFloat}
    @assert which_f[1] == "weibull pdf"
    @assert x > zero(T)
    T_k = CustomGaussQuadrature.materialize_scalar_spec_fn(T, which_f[3])
    @assert T_k > zero(T)
    log(T_k) + (T_k - one(T)) * log(x) - x^T_k
end
\end{lstlisting}
Since this weight function is a pdf, $\mu_0 = 1$. The Gauss rule is computed, via the Stieltjes procedure, using \texttt{stieltjes\_lnf\_new\_scr.jl}, which has inputs \texttt{which\_f}, \texttt{n}, the function \texttt{lnf\_typed\_fn} and \texttt{mu0}. The outputs include \texttt{nodes\_stieltjes}, \texttt{weights\_stieltjes} and \texttt{r} which is the number of nodes in the discrete approximation \eqref{eqn_discrete_approx_inner_prod_fns} chosen by the package. This computation is carried out using
\begin{lstlisting}[basicstyle=\ttfamily\small,breaklines=true,breakatwhitespace=false,columns=fullflexible]
which_f = ["weibull pdf", [0, Inf], "3.1"]	
n = 6

lnf_typed_fn = lnf_weibull_pdf
mu0 = 1

pkg_dir = dirname(dirname(pathof(CustomGaussQuadrature)))
include(joinpath(pkg_dir, "src", "stieltjes_lnf_new_scr.jl"))

nodes_stieltjes_ex2 = nodes_stieltjes
weights_stieltjes_ex2 = weights_stieltjes
println("nbits_stieltjes =", nbits_stieltjes)
println("r = ", r)
\end{lstlisting}
This code was also run with $n=6$ replaced by $n=14$.
%

%To compare the Gauss rules obtained using the Stieltjes procedure with that obtained using moment determinants, use:
%%
%\begin{lstlisting}[basicstyle=\ttfamily\small,breaklines=true,breakatwhitespace=false,columns=fullflexible]
%diff_nodes = nodes_stieltjes_ex2 - nodes_momentdets_ex2
%rel_diff_weights =
%    (weights_stieltjes_ex2 - weights_momentdets_ex2) ./ weights_momentdets_ex2
%
%println("maximum(abs.(nodes_stieltjes_ex2 - nodes_momentdets_ex2)) = ",
%    maximum(abs.(diff_nodes)))
%println("maximum(abs.((weights_stieltjes_ex2 - weights_momentdets_ex2) ./ weights_momentdets_ex2)) = ",
%    maximum(abs.(rel_diff_weights)))
%\end{lstlisting}
%%

\medskip

\noindent \textbf{Development and validation of the moment determinants implementation}

\smallskip

\noindent An earlier implementation of an adaptive-precision computation of a Gauss rule using moment determinants
was provided by the \textsf{R} package \textsf{custom.gauss.quad} (\citeauthor{Kabaila_custom.gauss.quad_2022},
\citeyear{Kabaila_custom.gauss.quad_2022}).
Although functional, this implementation was inefficient. The \textsf{Julia} package \textsf{CustomGaussQuadrature} was therefore developed to provide a much more efficient implementation making effective use of type-generic numerical programming.

The moment determinants implementations were extensively validated. For both \textsf{custom.gauss.quad} and \textsf{CustomGaussQuadrature}, Gauss rules associated with Legendre, Hermite and Generalized Laguerre polynomials were computed and compared with known results. For the non-classical \emph{Scaled chi pdf weight function}, the computations were also repeated at working precisions far above those selected adaptively, to check internal consistency. In addition, results from \textsf{CustomGaussQuadrature} were compared with those previously obtained using \textsf{custom.gauss.quad}.

\medskip

\noindent \textbf{Comparison of moment determinants and the Stieltjes procedure}

\smallskip

\noindent \textsf{Julia} uses just-in-time compilation, so the first execution of a computation may include both numerical work and compilation. Later executions in the same \textsf{Julia} session generally reuse the compiled code, provided that the argument types and code paths are unchanged. Each computation was therefore first run once with the same inputs to trigger compilation. The reported values are median wall-clock times from three subsequent runs and represent timings for an already-running \textsf{Julia} session. The computations were carried out on a laptop with a 13th Gen Intel Core i7-13700HX processor (2.10 GHz) and 32 GB of RAM.

To compare the Gauss rules obtained using the two methods, we use the nodes and weights, in their \texttt{Double64} forms. We report
\begin{equation*}
	\text{Maximum absolute node difference}
	=
	\max_i
	\left|
	\tau_i^{(\mathrm{S})}
	-
	\tau_i^{(\mathrm{M})}
	\right|
\end{equation*}
and
\begin{equation*}
		\text{Maximum absolute relative weight difference}
	=
	\max_i
	\left|
	\frac{
		\lambda_i^{(\mathrm{S})}
		-
		\lambda_i^{(\mathrm{M})}
	}{
		\lambda_i^{(\mathrm{M})}
	}
	\right|,
\end{equation*}
where the superscripts $\mathrm{S}$ and $\mathrm{M}$ refer to the
Stieltjes procedure and the moment determinants method, respectively.
%The Stieltjes procedure has a less stringent stopping criterion than the moment determinants method. 
%Because these two methods carry out Step 1 using
%entirely different procedures, 
%this comparison provides assurance 
%that the Gauss rule nodes and weights computed via moment determinants are sufficiently accurate.

\smallskip

\begin{table}[ht]
	\centering
	\caption{Moment determinants and Stieltjes procedure comparison for Example~1.}
	\label{tab:moment-stieltjes-example1}
\begin{tabular}{|l|c|}
	\hline
	\multicolumn{1}{|c|}{Quantity} & Value
	\\
	\hline	
	Moment determinants computation time (seconds)
	&
	 0.000728
	\\
	\hline
	Moment determinants \texttt{BigFloat} precision in bits
	&
	132
	\\
	\hline
	Stieltjes procedure computation time (seconds)
	&
	3.97
	\\
	\hline
	Stieltjes procedure \texttt{BigFloat} precision in bits
	&
	256
	\\
	\hline
	Stieltjes procedure number $r$ of auxiliary quadrature nodes
	&
	200
	\\
	\hline
	Maximum absolute node difference
	&
	$2.76 \times 10^{-20}$
	\\
	\hline
	Maximum absolute relative weight difference
	&
	$8.01 \times 10^{-19}$
	\\
	\hline
\end{tabular}
\end{table}

\smallskip

\begin{table}[ht]
	\centering
	\caption{Moment determinants and Stieltjes procedure comparison for Example~2 with $n = 6$.}
	\label{tab:moment-stieltjes-example2-n6}
\begin{tabular}{|l|c|}
	\hline
	\multicolumn{1}{|c|}{Quantity} & Value
	\\
	\hline	
	Moment determinants computation time (seconds)
	&
	0.000997
	\\
	\hline
	Moment determinants \texttt{BigFloat} precision in bits
	&
	132
	\\
	\hline
	Stieltjes procedure computation time (seconds)
	&
	81.3
	\\
	\hline
	Stieltjes procedure \texttt{BigFloat} precision in bits
	&
	256
	\\
	\hline
	Stieltjes procedure number $r$ of auxiliary quadrature nodes
	&
	650
	\\
	\hline
	Maximum absolute node difference
	&
	$3.63 \times 10^{-17}$
	\\
	\hline
	Maximum absolute relative weight difference
	&
	$3.56 \times 10^{-16}$
	\\
	\hline
\end{tabular}
\end{table}

\smallskip

\begin{table}[ht]
	\centering
	\caption{Moment determinants and Stieltjes procedure comparison for Example~2 with $n = 14$.}
	\label{tab:moment-stieltjes-example2-n14}
	\begin{tabular}{|l|c|}
		\hline
		\multicolumn{1}{|c|}{Quantity} & Value
		\\
		\hline	
		Moment determinants computation time (seconds)
		&
		 0.00737
		\\
		\hline
		Moment determinants \texttt{BigFloat} precision in bits
		&
		158
		\\
		\hline
		Stieltjes procedure computation time (seconds)
		&
		215
		\\
		\hline
		Stieltjes procedure \texttt{BigFloat} precision in bits
		&
		256
		\\
		\hline
		Stieltjes procedure number $r$ of auxiliary quadrature nodes
		&
		1029
		\\
		\hline
		Maximum absolute node difference
		&
		$2.95 \times 10^{-17}$
		\\
		\hline
		Maximum absolute relative weight difference
		&
		 $9.98 \times 10^{-16}$
		\\
		\hline
	\end{tabular}
\end{table}

\noindent In view of its more complex adaptive procedure and less stringent stopping tolerance, we expect the larger differences in Tables 2 and 3 to arise mainly from the Stieltjes procedure.
Increasing the number of Gauss rule nodes $n$ from 6 to 14 increases the working precision used by the moment determinants method and the number of auxiliary quadrature nodes used by the Stieltjes procedure. The computation times for both methods increase, although the moment determinants computation remains comparatively fast.

\newpage

\subsection{Difficulties in implementing the Stieltjes procedure}
\label{sect: Difficulties in implementing the Stieltjes procedure}

%At first sight, moment determinants seems to be the ``poor cousin'' of the Stieltjes procedure. Application of the moment determinants method requires a formula for the $s$'th moment $\mu_s$ and it is severely ill-conditioned, necessitating the use of high-precision arithmetic. The Stieltjes procedure requires only a formula for the weight function and so is more widely applicable. However, 

While the outline description of the Stieltjes procedure given in Subsection~\ref{sect_recursion_coeffs_stieltjes_procedure} is very simple, finding a suitable high-quality auxiliary quadrature for the computation of the inner products  
in \eqref{eqn:stieltjes procedure outline} can be difficult. For the \emph{Weibull pdf weight function} surprisingly large values of $r$ may need to be chosen, resulting in large computation times by comparison with the moment determinants method. 

For many weight functions, one needs to take account of the special nature of the weight function to avoid excessively large values of $r$. Consider, for example, a weight function that is an even function. In this case, see e.g. \cite{Chihara1978_IntroOrthogPolys}, $\alpha_0 = 0, ..., \alpha_{n-1}=0$ 
and  
\begin{equation*}
	\pi_{k+1}(x) = 	x \, \pi_k(x) - \beta_k \, \pi_{k-1}(x) \ \ \text{for} \ k = 0, 1, \dots, n-1.
\end{equation*}
It is convenient to compute 
the inner product \eqref{eqn_inner_prod_fns} by evaluating
\begin{equation*}
	\int_0^{\infty} \big(u(x) \, v(x) + u(-x) \, v(-x)\big) \, f(x) \, dx.
\end{equation*}
For the particular case that the weight function is not smooth at 0,
this split form is especially helpful computationally, since it avoids applying
a numerical quadrature rule across an interior point of nonsmoothness.
Consider, for example, the weight function $\exp(-|x|^{1/2})$ for $x \in \mathbb{R}$. 
Further examples of the need to take account of the special nature of the weight function 
when implementing the Stieltjes procedure are provided in subsection 2.2.4 of \cite{Gautschi2004}.

\section{Other Julia packages that use the Stieltjes procedure}

We briefly review other packages that compute custom-made Gauss rules using the Stieltjes procedure. 
The \textsf{PolyChaos} function \texttt{OrthoPoly} (with \texttt{addQuadrature} set to \texttt{true}) (\citeauthor{MuhlpfordtEtAl_PolyChaos}, \citeyear{MuhlpfordtEtAl_PolyChaos}) implements the initial transformation \eqref{eqn_initial_transf}, with the auxiliary quadrature rule either a Fej\'er quadrature (default) or  Clenshaw-Curtis quadrature. \texttt{Float64} arithmetic is used to compute the recursion coefficients in the three-term recurrence relation. The number of auxiliary quadrature nodes is $10 (n + 1)$ (default). For these default values, Tables 4, 5 and 6 compare the moment determinants method from \textsf{CustomGaussQuadrature} with \texttt{OrthoPoly}, with \texttt{addQuadrature} set to \texttt{true}, Fej\'er quadrature and this default number of auxiliary quadrature nodes.

\begin{table}[h]
	\centering
	\caption{Moment determinants and \texttt{OrthoPoly} comparison  for Example~1.}
	\label{tab:moment-orthopoly-example1}
	\begin{tabular}{|l|c|}
		\hline
		\multicolumn{1}{|c|}{Quantity} & Value
		\\
		\hline	
		Moment determinants computation time (seconds)
		&
	 0.000728
		\\
		\hline
		\texttt{OrthoPoly} computation time (seconds)
		&
		0.000073
		\\
		\hline
		\texttt{OrthoPoly} number of auxiliary quadrature nodes
		&
		40
		\\
		\hline
		Maximum absolute node difference
		&
		$0.0714$
		\\
		\hline
		Maximum absolute relative weight difference
		&
		$0.911$
		\\
		\hline
	\end{tabular}
\end{table}

\begin{table}[h]
	\centering
		\caption{Moment determinants and \texttt{OrthoPoly} comparison  for Example~2 with $n = 6$.}
	\label{tab:moment-orthopoly-example2-n6}
	\begin{tabular}{|l|c|}
		\hline
		\multicolumn{1}{|c|}{Quantity} & Value
		\\
		\hline	
		Moment determinants computation time (seconds)
		&
		0.000997
		\\
		\hline
		\texttt{OrthoPoly} computation time (seconds)
		&
		0.000048
		\\
		\hline
		\texttt{OrthoPoly} number of auxiliary quadrature nodes
		&
		70
		\\
		\hline
		Maximum absolute node difference
		&
		$5.03 \times 10^{-8}$ 
		\\
		\hline
		Maximum absolute relative weight difference
		&
		$2.89 \times 10^{-7}$ 
		\\
		\hline
	\end{tabular}
\end{table}

\begin{table}[h]
	\centering
	\caption{Moment determinants and \texttt{OrthoPoly} comparison  for Example~2 with $n = 14$.}
	\label{tab:moment-orthopoly-example2-n14}
	\begin{tabular}{|l|c|}
		\hline
		\multicolumn{1}{|c|}{Quantity} & Value
		\\
		\hline	
		Moment determinants computation time (seconds)
		&
		0.00737
		\\
		\hline
		\texttt{OrthoPoly} computation time (seconds)
		&
		0.000084
		\\
		\hline
		\texttt{OrthoPoly} number of auxiliary quadrature nodes
		&
		150
		\\
		\hline
		Maximum absolute node difference
		&
		$0.000147$ 
		\\
		\hline
		Maximum absolute relative weight difference
		&
		0.00339
		\\
		\hline
	\end{tabular}
\end{table}

\noindent Compare Tables 4, 5 and 6 with Tables 1, 2 and 3, respectively. The \texttt{OrthoPoly} default number of auxiliary quadrature nodes is at most one-fifth of the number of auxiliary quadrature nodes adaptively chosen for the Stieltjes procedure implemented in \textsf{CustomGaussQuadrature}. Also, the \texttt{Float64} arithmetic used by \texttt{OrthoPoly} has much lower precision than the adaptively chosen \texttt{BigFloat} arithmetic with 256-bit precision. Both factors contribute to the relatively large maximum absolute node and relative weight differences.

The \textsf{QuadGK} function \texttt{gauss} (\citeauthor{Johnson_QuadGK},
\citeyear{Johnson_QuadGK})
requires the support interval for the weight function to have finite endpoints. It approximates
the inner products in \eqref{eqn:stieltjes procedure outline} using adaptive Gauss-Kronrod quadrature. The observation at the end of subsection \ref{sect: Implement Stieltjes for Step 1} suggests that adaptive Gauss-Kronrod quadrature may be particularly suitable. 

\section{Discussion}

When formulae for the required moments are available, as is commonly the case in statistical applications, the moment determinants method can be used effectively despite its severe ill-conditioning, because the working precision is selected adaptively. We have implemented the Stieltjes procedure based on the ``general purpose'' initial transformation \eqref{eqn_initial_transf}. This procedure requires only a formula for the weight function and is therefore more widely applicable. However, to achieve a prescribed accuracy, this procedure requires adaptive selection of both the working precision and the number of auxiliary quadrature nodes. This makes the Stieltjes procedure more difficult to tune than the moment determinants method, for which only the working precision is adapted. Also, as illustrated by the \emph{Weibull pdf weight function}
with $n = 6$ and $n = 14$, the required number of auxiliary quadrature nodes can become very large, leading to long computation times.

Efficient use of the Stieltjes procedure may therefore require an implementation tailored to the special features of the weight function. For example, as noted in subsection \ref{sect: Difficulties in implementing the Stieltjes procedure},
an implementation for symmetric weight functions should use the fact that all the recursion coefficients $\alpha_k$'s are zero and compute the required inner products over the positive half-line using a symmetrized integrand. Such tailoring does not remove the need to adapt both the working precision and the number of auxiliary quadrature nodes, but it can greatly reduce the computational effort required.

\section*{Acknowledgements}

The core code for \textsf{CustomGaussQuadrature} is due solely to the author. Generative pre-trained LLMs were then used by the author to assist with code checking and code improvement for this package. 

\section*{Code availability}

 The \textsf{CustomGaussQuadrature} source code is
available at \url{https://github.com/pvkabaila/CustomGaussQuadrature.jl}; the
version described here is Version 3.1.0.

\section*{Appendix A: Details of the implementation of the moment determinants method  
	for Step 1 in code}

	We require that there is a
	formula for $\mu_s$, the $s$'th moment,
	for all nonnegative integers $s\le 2n-1$,
	in terms of mathematical functions that can be computed to arbitrary accuracy using \texttt{BigFloat} arithmetic.
	For the non-classical case, examples of probability distributions that satisfy this requirement, subject where necessary to parameter values for which moments through order $2n-1$ exist, are: 
	scaled chi, noncentral chi squared, truncated noncentral chi squared, truncated normal, generalized normal $\big(f(x) \propto  \exp\big(-(|x - \mu|/\alpha)^{\beta} \big), \: -\infty < x < \infty; \: \alpha>0, \: \beta > 0 \big)$, 
	Student's t, 
	truncated Student's t, generalized t as described on p.421 of \cite{JohnsonKotzBalakrishnan_ContUnivariate_vol2_2ndEd},
	noncentral t, F, inverse Gaussian, Weibull, inverse gamma, Pareto distributions of the first and second kind, exponentially-modified Gaussian
	(the distribution of the sum of independent normally and exponentially distributed random variables), Rice (pdf of the distance from the origin of a point with a circularly symmetric bivariate normal distribution), sine-power $\big(f(x) \propto \sin^m(\pi x), \: 0 < x < 1, m$ a nonnegative integer\big) and Kumaraswamy's double bounded $\big(f(x) \propto x^{a-1} \, (1 - x^a)^{b-1}, \: 0 < x < 1; \: a > 0, b > 0 \big)$ distributions.

	The script \texttt{gauss\_quad\_moment\_dets\_scr.jl} carries out both Steps 1 and 2, where Step 1 uses the moment determinants method. For the part of this script that carries out Step 1, we need to provide a formula for $\mu_k$, for all nonnegative integers $k\le 2n-1$,
	which can be evaluated to arbitrary accuracy using \texttt{BigFloat} arithmetic.
	This part consists of three components.
	The first component uses \texttt{BigFloat} arithmetic, with specified \texttt{b} bits of precision, to compute approximations to the following vectors in sequence: (1) $\big(\mu_0, \dots, \mu_{2n-1} \big)$, (2) 
	$\big(\Delta_0, \dots, \Delta_n\big)$ and 
	$\big(\Delta'_0, \dots, \Delta'_n\big)$ and (3) 
	$\big(\alpha_0, \alpha_1, \dots, \alpha_{n-1}\big)$ and 
	$\big(\beta_1, \beta_2, \dots, \beta_{n-1} \big)$.
	%, found using 
	%\eqref{eqn_alphak_moment_det}	
	%and
	%\eqref{eqn_betak_moment_det}, respectively. 
	The output of this component is 
	$\big(\alpha_0(\texttt{b}), \alpha_1(\texttt{b}), \dots, \alpha_{n-1}(\texttt{b})\big)$, 
	$\big(\beta_1(\texttt{b}), \beta_2(\texttt{b}), \dots,\beta_{n-1}(\texttt{b}) \big)$ and $\mu_0(\texttt{b})$,
	where we have made the dependence on \texttt{b} explicit in the notation.
	This component
	is a building block for the other two components.

	The second component steps through $\texttt{b} = 80, 106, 132, \dots, 392$ until the computation, by the first component, of  $\big(\alpha_0(\texttt{b}), \alpha_1(\texttt{b}), \dots, \alpha_{n-1}(\texttt{b})\big)$ and 
	$\big(\beta_1(\texttt{b}), \beta_2(\texttt{b}), \dots,$ $\beta_{n-1}(\texttt{b}) \big)$ does not fail and all elements of the latter vector are positive. The resulting value of \texttt{b} is denoted $\texttt{b}_1$.
	The output of the second component is $\big(\alpha_0(\texttt{b}_1), \alpha_1(\texttt{b}_1), \dots,$ $\alpha_{n-1}(\texttt{b}_1)\big)$,
	$\big(\beta_1(\texttt{b}_1), \beta_2(\texttt{b}_1), \dots,\beta_{n-1}(\texttt{b}_1) \big)$ and $\mu_0(\texttt{b}_1)$. Of course, there is no guarantee that 
	this output
	provides sufficiently good approximations for the successful completion of Step 1.

	The third component steps through $\texttt{b} = \texttt{b}_1, \texttt{b}_2, \texttt{b}_3, \dots$,
	where $\texttt{b}_j = \texttt{b}_1 + 26 \, (j-1)$ and $\texttt{b}_j \le 444$, and computes
	\begin{align*}
		\big(a_1(\texttt{b}), a_2(\texttt{b}), \dots, a_n(\texttt{b})\big)
		&=\big(\alpha_0(\texttt{b}), \alpha_1(\texttt{b}), \dots, \alpha_{n-1}(\texttt{b})\big), 
		\\
		\big(b_1(\texttt{b}), b_2(\texttt{b}), \dots,b_{n-1}(\texttt{b}) \big)
		&= \big(\sqrt{\beta_1(\texttt{b})}, \sqrt{\beta_2(\texttt{b})}, \dots,\sqrt{\beta_{n-1}(\texttt{b})} \big)  	
	\end{align*}
	and $\mu_0(\texttt{b})$.
	To assure ourselves that the chosen number of bits $\texttt{b}$
	of precision is sufficiently large, we use
	the simple method described in Section \ref{sect_meth_moment_dets_step1_julia}.
	Accordingly, the third component  computes 
	\begin{align}
		\label{eqn_max_est_abs_error_as}
		\max_{i=1, \dots, n} \big|a_i(\texttt{b}_j) - a_i(\texttt{b}_{j+1})\big|	
		\\
		\label{eqn_max_est_rel_error_bs}
		\max_{i=1, \dots, n-1} \frac{\big|b_i(\texttt{b}_j) - b_i(\texttt{b}_{j+1})\big|}{b_i(\texttt{b}_{j+1})},
	\end{align}
	for $j=1, 2, \dots$, and sets 
	$\texttt{b}_{\rm final} = \texttt{b}_{j+1}$ 
	when both \eqref{eqn_max_est_abs_error_as}
	and \eqref{eqn_max_est_rel_error_bs} are less than 
	$10^{-18}$. The output of Step 1 is
	\begin{equation*}
		\big(a_1(\texttt{b}_{\rm final}), a_2(\texttt{b}_{\rm final}), \dots, a_n(\texttt{b}_{\rm final})\big),
	\end{equation*}
	\begin{equation*}
		\big(b_1(\texttt{b}_{\rm final}), b_2(\texttt{b}_{\rm final}), \dots, b_{n-1}(\texttt{b}_{\rm final})\big),
	\end{equation*}
	and the scalars $\mu_0(\texttt{b}_{\rm final})$ and $\texttt{b}_{\rm final}$.

	\section*{Appendix B: The method used to compute the discrete approximation 	\eqref{eqn_discrete_approx_inner_prod_fns}
	to $\boldsymbol{(u, v)}$}
	\label{sect:Method used to compute discrete approximation}

	A discrete approximation to the right-hand side of 	\eqref{eqn_initial_transf} can be found as follows.
	Let 
	$h(y) = g\big(\varphi(y)\big) \, f\big(\varphi(y)\big) \, \varphi'(y)$, where
	\begin{equation*}
		\varphi'(y) = 
		\begin{cases}
			(1/2) (b - a) &\text{if } -\infty < a < b < \infty
			\\
			2 / (1 + y)^2 &\text{if } -\infty = a < b < \infty
			\\
			2 / (1 - y)^2 &\text{if } -\infty < a < b = \infty
			\\
			(1 + y^2) / \big(1 - y^2 \big)^2  &\text{if } -\infty = a < b = \infty.
		\end{cases}	
	\end{equation*}
	A high-quality quadrature rule then provides a discrete approximation to the integral $\int_{-1}^1 h(y) \, dy$.
	\cite{Gautschi2004} used a Fej\'er quadrature rule.
	Instead, we use Gauss--Legendre quadrature with $r$ nodes to approximate
	\begin{equation*}
		\int_{-1}^1 h(y) \, dy \quad \text{by} \quad \sum_{i =1}^r \xi_i \, h(y_i),	
	\end{equation*}
	where $y_1, \dots, y_r$ are the nodes and $\xi_1, \dots, \xi_r$ are the corresponding weights.
	The method used to choose both $r$ and the working precision is described at the end of this appendix. Now 
	\begin{equation*}
		%	\label{eqn_discrete_approx_inner_prod_fns}
		\sum_{i =1}^r \xi_i \, h(y_i) = \sum_{i =1}^r w_i \, g(x_i),	
	\end{equation*}
	where $w_i =  \xi_i \, \varphi'(y_i) \, f\big(\varphi(y_i)\big)$
	and $x_i = \varphi(y_i)$. To summarize, we 
	approximate the inner product $(u, v)$ by
	\eqref{eqn_discrete_approx_inner_prod_fns}.
	All that the user needs to provide is a \textsf{Julia}
	function to evaluate $f$.

	Consider the case that either $-\infty = a$ or $b = \infty$, or both.
	It was found that the computation of the $w_i$'s using 
	\texttt{Double64} arithmetic, provided by \textsf{DoubleFloats},
	could result in some of these being \texttt{NaN}'s. To solve this problem, we compute the $\log(w_i)$'s instead using
	\begin{equation*}
		\log(w_i) = \log(\xi_i) + \log \big(\varphi'(y_i) \big) + \log\big(f\big(\varphi(y_i)\big) \big),
	\end{equation*}
	where
	\begin{equation*}
		\log\big(\varphi'(y)\big) = 
		\begin{cases}
			\log(2) - 2 \log(1 + y) &\text{if } -\infty = a < b < \infty
			\\
			\log(2) - 2 \log(1 - y) &\text{if } -\infty < a < b = \infty
			\\
			\log(1 + y^2) - 2 \log \big(1 - y^2 \big)  &\text{if } -\infty = a < b = \infty.
		\end{cases}	
	\end{equation*}
	All that the user needs to provide is a \textsf{Julia}
	function to evaluate $\log(f)$. Since $g = uv$, 
	\begin{align*}
		w_i \, g(x_i)
		&= w_i \, u(x_i) \, v(x_i)
		\\
		&= \text{sign}\big(u(x_i)\big) \, \text{sign}\big(v(x_i)\big) \,
		w_i \, \big|u(x_i)\big| \, \big|v(x_i)\big|
		\\
		&= \text{sign}\big(u(x_i)\big) \, \text{sign}\big(v(x_i)\big) \,
		\exp\Big(\log(w_i) + \log\big(\big|u(x_i)\big|\big) + \log\big(\big|v(x_i)\big|\big) \Big),
	\end{align*}
	which is used to compute \eqref{eqn_discrete_approx_inner_prod_fns}.

	\subsection*{Choice of the number $\boldsymbol{r}$ of auxiliary quadrature nodes and working precision}
	
	The value of $r$ and the working precision are not fixed in advance. 
	Let \texttt{b} denote the number of bits of
	\texttt{BigFloat} precision and set $\texttt{b}=128$ for 
	\texttt{Double64} precision. For any given candidate values of $r$ and \texttt{b}, let
	\begin{align}
		\label{eqn: computed values of the ai and bi for given r and b}
		\begin{split}
			\big(a_1(r, \texttt{b}), a_2(r, \texttt{b}), \dots, a_n(r, \texttt{b})\big)
			&=\big(\alpha_0(r, \texttt{b}), \alpha_1(r, \texttt{b}), \dots, \alpha_{n-1}(r, \texttt{b})\big), 
			\\
			\big(b_1(r, \texttt{b}), b_2(r, \texttt{b}), \dots,b_{n-1}(r, \texttt{b}) \big)
			&= \big(\sqrt{\beta_1(r, \texttt{b})}, 
			\sqrt{\beta_2(r, \texttt{b})}, \dots,\sqrt{\beta_{n-1}(r, \texttt{b})} \big)
		\end{split}
	\end{align}
	denote the vectors computed by applying the Stieltjes procedure 
	\eqref{eqn:stieltjes procedure outline}
	with the discrete approximation
	\eqref{eqn_discrete_approx_inner_prod_fns}.
	For any given candidate value $r$ and $\texttt{b}_2 \ge \texttt{b}_1 + 32$, we say that
	the computation of the vectors 
	\eqref{eqn: computed values of the ai and bi for given r and b}
	with precision $\texttt{b}_1$ agrees with the computation of these vectors
	with precision
	$\texttt{b}_2$ when
	\begin{equation*}
		\max_i |a_i(r, \texttt{b}_1)-a_i(r, \texttt{b}_2)| < 10^{-22}
		\quad \text{and} \quad
		\max_i \left|\frac{(b_i(r, \texttt{b}_1)-b_i(r, \texttt{b}_2))}{b_i(r, \texttt{b}_2)}\right| < 10^{-22}.
	\end{equation*}

	We consider the sequence of candidate values of $r$:
	\begin{equation}
		\label{eqn:sequence of rk}	
		r_j = j \, (n + \texttt{offset}), \qquad j = 3, 4, 5, \dots, j_{\max},
	\end{equation}
	where the default values of \texttt{offset} and $j_{\max}$ are $7$ and $80$,
	respectively.
	For each $r_j$ the algorithm first computes these vectors for
	$\texttt{b}_2=256$ and $\texttt{b}_1=128$. If the computation for
	$\texttt{b}_1=128$ contains no \texttt{NaN}'s and there is agreement in the
	sense above, then we set $\texttt{b}_{\rm final}(r_j)=256$.
	If this first check does not succeed, then the algorithm computes these
	vectors for $\texttt{b}_1=224$ and compares them with those already computed
	for $\texttt{b}_2=256$. If there is agreement in the sense above, then we set
	$\texttt{b}_{\rm final}(r_j)=256$.
	Otherwise the algorithm computes these vectors for $\texttt{b}_2=512$ and
	compares them with the earlier vectors for $\texttt{b}_1=256$. If there is
	agreement in this comparison, then we set $\texttt{b}_{\rm final}(r_j)=512$.
	If not, then the algorithm computes these vectors for $\texttt{b}_1=480$ and
	compares them with those already computed for $\texttt{b}_2=512$. If there is
	agreement in this final comparison, then we set $\texttt{b}_{\rm final}(r_j)=512$.
	Otherwise the corresponding problem is rejected rather than returned with
	coefficients of uncertain accuracy. Thus, for each candidate value $r_j$, the
	procedure either selects a working precision $\texttt{b}_{\rm final}(r_j) \in
	\{256,512\}$ and returns the corresponding vectors 
	\eqref{eqn: computed values of the ai and bi for given r and b}, or rejects the problem.
	For an accepted candidate value $r_j$, define
	$a_i^{(j)} = a_i(r_j, \texttt{b}_{\rm final}(r_j))$ and
	$b_i^{(j)} = b_i(r_j, \texttt{b}_{\rm final}(r_j))$.

	Let $r_{\rm final}$ denote $r_{j+1}$ for the smallest value of $j$ such that
	\begin{equation*}
		\max_i \big|a_i^{(j+1)} - a_i^{(j)}\big| \le 10^{-17}
		\qquad \text{and} \qquad
		\max_i \left|
		\frac{b_i^{(j+1)} - b_i^{(j)}}{b_i^{(j+1)}}
		\right| \le 10^{-17 }.
	\end{equation*}
	Thus, $10^{-17}$ is a stopping tolerance applied to differences between successive computations. 
	These differences are used as practical error indicators, but they are not guaranteed upper bounds on the errors in the returned recursion coefficients.
	The output is therefore 
	$r_{\rm final}$, the working precision
	$\texttt{b}_{\rm final}(r_{\rm final})$, and the corresponding vectors
	\eqref{eqn: computed values of the ai and bi for given r and b}.

%\section{Bibliography}

%\bibliographystyle{authordate1}
%\bibliographystyle{elsevier}

% Specify a relative path, going 6 directories
% upward and then into the BIBLIOGRAPHY folder.
%\bibliography{../../../../../../BIBLIOGRAPHY/Bibliography}

\begin{thebibliography}{}
	
	\bibitem[\protect\citename{Abeysekera \& Kabaila, }2017]{AbeysekeraKabaila2017}
	Abeysekera, W., \& Kabaila, P. 2017.
	\newblock Optimized recentered confidence spheres for the multivariate normal
	mean.
	\newblock {\em Electronic Journal of Statistics}, {\bf 11}, 1798--1826.
	
	\bibitem[\protect\citename{Chihara, }1978]{Chihara1978_IntroOrthogPolys}
	Chihara, T.S. 1978.
	\newblock {\em An Introduction to Orthogonal Polynomials}.
	\newblock Mathematics and Its Applications, vol. 13.
	\newblock New York: Gordon and Breach.
	
	\bibitem[\protect\citename{Dunnett, }1989]{Dunnett}
	Dunnett, C.W. 1989.
	\newblock Algorithm {AS} 251: Multivariate normal probability integrals with
	product correlation structure.
	\newblock {\em Journal of the Royal Statistical Society, Series C (Applied
		Statistics)}, {\bf 38}, 564--579.
	
	\bibitem[\protect\citename{Dunnett \& Sobel, }1955]{DunnettSobel}
	Dunnett, C.W., \& Sobel, M. 1955.
	\newblock Approximations to the probability integral and certain percentage
	points of a multivariate analogue of {S}tudent's t-distribution.
	\newblock {\em Biometrika}, {\bf 42}, 258--260.
	
	\bibitem[\protect\citename{Gautschi, }1968]{Gautschi1968}
	Gautschi, W. 1968.
	\newblock Construction of {G}auss-{C}hristoffel quadrature formulas.
	\newblock {\em Mathematics of Computation}, {\bf 22}, 251--270.
	
	\bibitem[\protect\citename{Gautschi, }1983]{Gautschi1983}
	Gautschi, W. 1983.
	\newblock How and how not to check {G}aussian quadrature formulae.
	\newblock {\em BIT}, {\bf 23}, 209--216.
	
	\bibitem[\protect\citename{Gautschi, }1994]{Gautschi1994}
	Gautschi, W. 1994.
	\newblock Algorithm 726: {ORTHPOL} - {A} package of routines for generating
	orthogonal polynomials and {G}auss-type quadrature rules.
	\newblock {\em ACM Transactions on Mathematical Software}, {\bf 20}, 21--62.
	
	\bibitem[\protect\citename{Gautschi, }2004]{Gautschi2004}
	Gautschi, W. 2004.
	\newblock {\em Orthogonal Polynomials: Computation and Approximation}.
	\newblock New York: Oxford University Press.
	
	\bibitem[\protect\citename{Genz \& Bretz, }2009]{GenzBretz}
	Genz, A., \& Bretz, F. 2009.
	\newblock {\em Computation of Multivariate Normal and t Probabilities}.
	\newblock London: Springer.
	
	\bibitem[\protect\citename{Gupta \& Panchapakesan,
	}2002]{GuptaPanchapakesan2002}
	Gupta, S.S., \& Panchapakesan, S. 2002.
	\newblock {\em Multiple Decision Procedures: Theory and Methodology of
		Selecting and Ranking Populations}.
	\newblock Philadelphia: SIAM.
	
	\bibitem[\protect\citename{Hochberg \& Tamhane, }1987]{HochbergTamhane1987}
	Hochberg, Y., \& Tamhane, A.C. 1987.
	\newblock {\em Multiple Comparison Procedures}.
	\newblock New York: Wiley.
	
	\bibitem[\protect\citename{Johnson {\em et~al.},
	}1995]{JohnsonKotzBalakrishnan_ContUnivariate_vol2_2ndEd}
	Johnson, N.L., Kotz, S., \& Balakrishnan, N. 1995.
	\newblock {\em Continuous Univariate Distributions}. second edn.
	\newblock  Vol. 2.
	\newblock New York: John Wiley.
	
	\bibitem[\protect\citename{Johnson, }2026]{Johnson_QuadGK}
	Johnson, Steven~G. 2026.
	\newblock {\em {QuadGK.jl}: {G}auss--{K}ronrod integration in {J}ulia}.
	\newblock \url{https://github.com/JuliaMath/QuadGK.jl}.
	\newblock {V}ersion 2.11.3.
	
	\bibitem[\protect\citename{Kabaila, }2022]{Kabaila_custom.gauss.quad_2022}
	Kabaila, P. 2022.
	\newblock {\em {custom.gauss.quad}: R package for custom-made {G}auss
		quadrature}.
	\newblock \url{https://CRAN.R-project.org/package=custom.gauss.quad}.
	\newblock {V}ersion 1.0.0.
	
	\bibitem[\protect\citename{Kabaila, }2026]{Kabaila_CustomGaussQuadrature_2026}
	Kabaila, P. 2026.
	\newblock {\em {CustomGaussQuadrature.jl}: Julia package for custom-made
		{G}auss quadrature}.
	\newblock \url{https://github.com/pvkabaila/CustomGaussQuadrature.jl}.
	\newblock {V}ersion 3.1.0.
	
	\bibitem[\protect\citename{Kabaila \& Giri, }2009]{KabailaGiri2009JSPI}
	Kabaila, P., \& Giri, K. 2009.
	\newblock Confidence intervals in regression utilizing prior information.
	\newblock {\em Journal of Statistical Planning and Inference}, {\bf 139},
	3419--3429.
	
	\bibitem[\protect\citename{Kabaila \& Giri, }2013]{KabailaGiri2013}
	Kabaila, P., \& Giri, K. 2013.
	\newblock Further properties of frequentist confidence intervals in regression
	that utilize uncertain prior information.
	\newblock {\em Australian \& New Zealand Journal of Statistics}, {\bf 55},
	259--270.
	
	\bibitem[\protect\citename{Kabaila \& Perera, }2023]{KabailaPerera2023}
	Kabaila, P., \& Perera, A. 2023.
	\newblock Model averaged tail area confidence intervals in nested linear
	regression models.
	\newblock {\em Australian \& New Zealand Journal of Statistics}, {\bf 65},
	364--378.
	
	\bibitem[\protect\citename{Kabaila \& Ranathunga, }2021]{KabailaRanathunga2021}
	Kabaila, P., \& Ranathunga, N. 2021.
	\newblock Computation of the expected value of a chi-distributed random
	variable.
	\newblock {\em Computational Statistics}, {\bf 36}, 313--332.
	
	\bibitem[\protect\citename{Kabaila \& Tissera, }2014]{KabailaTissera2014}
	Kabaila, P., \& Tissera, D. 2014.
	\newblock Confidence intervals in regression that utilize uncertain prior
	information about a vector parameter.
	\newblock {\em Australian \& New Zealand Journal of Statistics}, {\bf 56},
	371--383.
	
	\bibitem[\protect\citename{Kabaila {\em et~al.},
	}2016]{KabailaWelshAbeysekera2016}
	Kabaila, P., Welsh, A.H., \& Abeysekera, W. 2016.
	\newblock Model-averaged confidence intervals.
	\newblock {\em Scandinavian Journal of Statistics}, {\bf 43}, 35--48.
	
	\bibitem[\protect\citename{Kabaila {\em et~al.},
	}2020]{KabailaWelshWijethunga2020_JSPI}
	Kabaila, P., Welsh, A.H., \& Wijethunga, C. 2020.
	\newblock Finite sample properties of confidence intervals centered on a model
	averaged estimator.
	\newblock {\em Journal of Statistical Planning and Inference}, {\bf 207},
	10--26.
	
	\bibitem[\protect\citename{Miller, }1981]{Miller1981}
	Miller, R.G. 1981.
	\newblock {\em Simultaneous Statistical Inference}. 2nd edn.
	\newblock New York: Springer.
	
	\bibitem[\protect\citename{Miwa {\em et~al.}, }2003]{MiwaHayterKuriki2003}
	Miwa, T., Hayter, A.J., \& Kuriki, S. 2003.
	\newblock The evaluation of general non-centred orthonant probabilities.
	\newblock {\em Journal of the Royal Statistical Society, Series B}, {\bf 65},
	223--234.
	
	\bibitem[\protect\citename{Mori, }1988]{Mori1988}
	Mori, M. 1988.
	\newblock The double exponential formula for numerical integration over the
	half infinite interval.
	\newblock {\em Pages  367--379 of:} Agarwal, R.P., Chow, Y.M., \& Wilson, S.J.
	(eds), {\em Numerical Mathematics (Singapore 1988)}.
	\newblock Basel: Birkhauser.
	
	\bibitem[\protect\citename{{M{\"u}hlpfordt} {\em et~al.},
	}2026]{MuhlpfordtEtAl_PolyChaos}
	{M{\"u}hlpfordt}, Tillmann, {Zahn}, Frederik, {Hagenmeyer}, Veit, \&
	{Faulwasser}, Timm. 2026.
	\newblock {\em {PolyChaos.jl}: Orthogonal Polynomials, Quadrature, and
		Polynomial Chaos}.
	\newblock \url{https://github.com/SciML/PolyChaos.jl}.
	\newblock {V}ersion 1.2.0.
	
	\bibitem[\protect\citename{Stewart,
	}2001]{Stewart_MatrixAlgorithms_Eigensystems_2001}
	Stewart, G.W. 2001.
	\newblock {\em Matrix Algorithms. Volume II: Eigensystems}.
	\newblock Philadelphia: Society for Industrial and Applied Mathematics.
	
\end{thebibliography}

\end{document}